\documentstyle[12pt,a4]{article}
\input epsf

\global\arraycolsep=2pt

\begin{document}

\begin{titlepage}

\begin{flushright}
CERN-TH.7312/94\\
SHEP 93/94-25 \\
hep-ph/9407394
\end{flushright}

\vspace{0.3cm}

\begin{center}
\Large\bf Cancellation of Renormalon Ambiguities in the\\
Heavy Quark Effective Theory
\end{center}

\vspace{0.8cm}

\begin{center}
Matthias Neubert\\
{\sl Theory Division, CERN, CH-1211 Geneva 23, Switzerland}\\
\vspace{0.3cm}
and\\
\vspace{0.3cm}
Chris T. Sachrajda\\
{\sl Department of Physics, University of Southampton\\
Southampton SO17 1BJ, United Kingdom}
\end{center}

\vspace{0.8cm}

\begin{abstract}
Recently, it has been shown that the concept of the pole mass of a
heavy quark becomes ambiguous beyond perturbation theory, because of
the presence of infrared renormalons. We argue that the predictions
of heavy quark effective theory, whose construction is based on the
pole mass, are free of such ambiguities. In the $1/m_Q$ expansion of
physical quantities, infrared and ultraviolet renormalons compensate
each other between coefficient functions and matrix elements. We
trace the appearance of these compensations for current-induced
exclusive heavy-to-heavy and heavy-to-light transitions, and for
inclusive decays of heavy hadrons. In particular, we show that the
structure of the heavy quark expansion is not obscured by
renormalons, and none of the predictions of heavy quark effective
theory are invalidated.
\end{abstract}

\centerline{(submitted to Nuclear Physics B)}
\bigskip\bigskip

\noindent
CERN-TH.7312/94\\
July 1994

\end{titlepage}

\section{Introduction}

For heavy quarks, the non-relativistic bound-state picture suggests
the notion of the pole mass $m_Q^{\rm pole}$ defined, order by order
in perturbation theory, as the position of the singularity in the
renormalized quark propagator. The pole mass is gauge invariant,
infrared finite, and renormalization-scheme independent \cite{Tarr}.
In the context of perturbation theory, it is thus a meaningful
``physical'' parameter. Once non-perturbative effects are taken into
account, however, this concept needs to be generalized, since in
reality there is no pole in the quark propagator because of
confinement. Recently, it has been shown that signals for such
non-perturbative effects can be found in the asymptotic behaviour of
perturbation theory itself. The presence of infrared renormalons in
the perturbative series that relates the pole mass to a mass defined
at short distances leads to an unavoidable ambiguity of order
$\Lambda_{\rm QCD}$ in the definition of $m_Q^{\rm pole}$
\cite{BBren,Bigiren}. The appearance of renormalons signals that
perturbation theory is incomplete without the inclusion of
non-perturbative corrections. In fact, much of the non-perturbative
structure of a theory can be inferred from a study of the
singularities of correlation functions after Borel transformation
with respect to the coupling constant. The application of this
approach to QCD was pioneered by 't Hooft \cite{tHof}. The positions
of the singularities on the positive real axis signal the magnitude
of non-perturbative corrections. In turn, the structure of
non-perturbative corrections implies constraints for the structure of
infrared renormalons \cite{Pari}--\cite{Bene}.

The existence of a ``physical'' definition of the mass of a heavy
quark, which agrees with the pole mass up to terms of order
$\Lambda_{\rm QCD}$, plays a crucial role in the construction of the
heavy quark effective theory (HQET) \cite{EiHi}--\cite{review}, which
by now has become the main theoretical tool used to analyze the
properties and decays of hadrons containing a heavy quark. In view of
the intrinsic ambiguity in the definition of the pole mass, the
question arises whether the HQET is an inconsistent effective theory,
whose predictions are plagued by renormalon ambiguities. The purpose
of this paper is to demonstrate that this is not the case.
Renormalons enter HQET predictions because one tries to separate
perturbative and non-perturbative (as opposed to short- and
long-distance) effects into coefficient functions and matrix
elements. Infrared renormalons appear in the coefficient functions
since soft loop momenta give a non-negligible contribution to the
Feynman integrals which appear in their calculation. Similarly,
ultraviolet renormalons appear in the matrix elements because of the
power divergence of Feynman integrals in the HQET. In this paper, we
argue that in predictions for physical quantities such as weak decay
amplitudes, renormalon ambiguities cancel between coefficient
functions and matrix elements.\footnote{For the case of the heavy
quark two-point function this cancellation has been demonstrated in
Ref.~\cite{BBren}.}
A generic HQET prediction for a physical quantity $A(m_Q)$ is of the
form
\begin{equation}
   A(m_Q) = C_0(m_Q/\mu)\,M_0(\mu)
   + {1\over m_Q}\,C_1(m_Q/\mu)\,M_1(\mu) + \ldots \,.
\end{equation}
We will trace the cancellation of renormalons explicitly to
subleading order in $1/m_Q$, by showing that the infrared renormalon
in $C_0$ cancels against an ultraviolet renormalon in the matrix
element $M_1$, so that the sum of the two terms on the right-hand
side is unambiguous. In more complicated processes such as
flavour-changing transitions between two heavy hadrons of different
velocity, the way in which these cancellations take place is rather
non-trivial. However, that they take place should not be a surprise.
In fact, the appearance of renormalons could be avoided if in the
construction of the HQET one would follow the idea of Wilson's
operator product expansion (OPE) \cite{Wils} literally
\cite{BBren,Bigiren}. The OPE is not designed to separate
perturbative and non-perturbative effects, but to disentangle the
physics on different distance scales. This is not accomplished when
one uses dimensional regularization in the calculation of the
coefficient functions. Instead, one should introduce a hard
factorization scale $\mu<m_Q$ by cutting out momenta $k<\mu$ from the
Feynman diagrams which determine the Wilson coefficients, and
attribute these contributions to the matrix elements. In practice,
this procedure is impracticable and awkward, but it would eliminate
the infrared renormalons from the coefficient functions and the
ultraviolet renormalons from the matrix elements. What is important
is that in the HQET such a program could be implemented without
changing the transformation property of the effective Lagrangian
under the spin-flavour symmetry \cite{Agli}. Hence, the structure of
the predictions obtained using the HQET remains unaffected. This
implies that renormalons enter the usual (practical) form of the HQET
in such a way that they do not spoil the relations imposed by heavy
quark symmetry \cite{Isgu} and the equation of motion (including the
vanishing of certain $1/m_Q$ corrections at zero recoil \cite{Luke}),
and they do not increase the number of hadronic form factors that
appear in a given order of the $1/m_Q$ expansion.

Let us note in passing that in the lattice formulation of the HQET
(or indeed in any regularization scheme with a dimensionful cut-off)
one encounters ultraviolet divergences which behave as powers of the
ultraviolet cutoff (i.e.\ inverse powers of the lattice spacing).
These power divergences are due to the mixing of higher dimensional
operators with lower dimensional ones. They become more severe as
higher-order terms in the $1/m_Q$ expansion are calculated. The
presence of power divergences, and the fact that they are likely to
imply the existence of non-perturbative effects, and hence to require
non-perturbative subtractions, was explained in Ref.~\cite{mms}. The
close connection between the presence of power divergences and that
of ultraviolet renormalons in matrix elements of higher dimensional
operators in the HQET was pointed out in Ref.~\cite{BBren}, and will
become apparent below. Techniques for the non-perturbative
subtraction of the power divergences in lattice simulations are being
developed \cite{ms}; a brief outline of the approach can be found in
Ref.~\cite{dallas}.

The outline of the paper is as follows: In Sect.~\ref{sec:2}, we
briefly discuss the appearance of renormalons in the asymptotics of
perturbation theory and their relation to singularities in the Borel
transform of correlation functions. To obtain a renormalon calculus
which is convenient for explicit calculations, we follow
Refs.~\cite{BBren,Bene} and consider QCD in the limit of a large
number of light quark flavours. We then recall some of the reasoning
behind the usual construction of the HQET and show in which way it is
affected by infrared renormalons in the pole mass of the heavy quark.
In Sects.~\ref{sec:3} and \ref{sec:4}, we study the cancellation of
renormalons in exclusive heavy-to-heavy and heavy-to-light decay
processes. In the first type of decays, the symmetries of the
effective theory imply a set of non-trivial consistency conditions,
which relate the infrared renormalons in the coefficient functions of
bilinear heavy quark currents to the infrared renormalon in the pole
mass. We derive the exact form of these relations, which are
independent of any unknown hadronic matrix element. We then check
them to order $1/N_f$. We also show with an explicit calculation that
a sum rule recently derived by Shifman et al.\ \cite{Bigisum}, which
has been used to put a bound on the hadronic form factor that enters
the extraction of $|\,V_{cb}|$ from semileptonic decays, cannot be
correct, due to a mismatch of infrared and ultraviolet renormalons.
In Sect.~\ref{sec:5}, we show that renormalon contributions cancel in
inclusive, current-induced decays of hadrons containing a heavy
quark. This proves a conjecture of Bigi et al.\ \cite{Bigiren},
although we do not agree on the details of the cancellation. In
Sect.~\ref{sec:6}, we summarize our results and give some
conclusions.

\section{Renormalons and the Construction of the HQET}
\label{sec:2}

Given a perturbative series for some quantity $F(\alpha_s)$ in terms
of the coupling constant $\alpha_s(\mu)$ renormalized at some scale
$\mu$,
\begin{equation}\label{Fseries}
   F(\alpha_s) = \sum_{n=0}^\infty F_n\,\bigg(
   {\beta_0\over 4\pi}\,\alpha_s(\mu) \bigg)^n \,,
\end{equation}
where $\beta_0=11-\frac{2}{3} N_f$ is the first coefficient of the
$\beta$-function, we define the Borel transform $\widetilde F(u)$ of
$F(\alpha_s)$ by
\begin{equation}
   \widetilde F(u) = F_0\,\delta(u) + \sum_{n=0}^\infty {1\over n!}\,
   F_{n+1}\,u^n \,.
\end{equation}
If the series is Borel summable, the function $F(\alpha_s)$ can be
reconstructed from its Borel transform using the integral relation
\begin{equation}\label{BTinv}
   F(\alpha_s) = \int\limits_0^\infty\!{\rm d}u\,
   \exp\bigg( -{4\pi u\over\beta_0\,\alpha_s(\mu)} \bigg)\,
   \widetilde F(u) \,.
\end{equation}
However, if the coefficients $F_n$ in (\ref{Fseries}) develop a
factorial divergence for large $n$, the Borel transform $\widetilde
F(u)$ can have singularities on the integration contour, and the
na\"\i ve Borel summation fails. In such a case, the result of the
integration depends on a regularization (or resummation)
prescription, and $F(\alpha_s)$ is not uniquely defined in terms of
$\widetilde F(u)$.

In QCD, one source of divergence in the expansion coefficients of a
perturbative series is related to higher-order diagrams in which a
virtual gluon line with momentum $k$ is dressed by a number of
fermion, gluon and ghost loops.\footnote{We hasten to add that in a
non-abelian theory bubble summation is not a gauge-invariant
procedure. This is one of the reasons why we will have to use a
large-$N_f$ expansion to obtain a consistent renormalon calculus, see
below.}
Effectively, this introduces the running coupling constant $g_s(k)$
at the vertices. Since the coupling constant increases for low
momenta because of asymptotic freedom, the insertion of additional
bubbles drives the gluon line to increasingly softer momentum, i.e.\
the infrared region in Feynman integrals becomes more
important. When the running coupling constant is expressed in terms
of a fixed coupling constant renormalized at some large scale $\mu$,
using
\begin{equation}
   \alpha_s(k) \simeq {\alpha_s(\mu)\over\displaystyle 1 -
    {\beta_0\over 4\pi}\,\alpha_s(\mu)\,\ln{\mu^2\over k^2}}
    = \sum_{n=0}^\infty \big[ \alpha_s(\mu) \big]^{n+1}
    \bigg( {\beta_0\over 4\pi}\,\ln{\mu^2\over k^2} \bigg)^n \,,
\end{equation}
the appearance of powers of large logarithms leads to a factorial
divergence in the expansion coefficients $F_n$ in (\ref{Fseries}).
Associated with this are renormalon singularities in the Borel
transform $\widetilde F(u)$.

In our case, the renormalon singularities will occur as single poles
on the real axis in the Borel plane. Poles on the positive real axis,
which arise from the low-momentum region of Feynman diagrams, are
called infrared renormalons.\footnote{Similarly, poles on the
negative real axis arise from the high-momentum region and are called
ultraviolet renormalons.}
Let us denote the positions of these poles by $u_i$ and their
residues by $r_i$, so that
\begin{equation}
   \widetilde F(u) = \sum_i {r_i\over u-u_i} + \ldots \,,
\end{equation}
where the ellipses represent terms that are regular for $u>0$. For
the calculation of the inverse Borel transform from (\ref{BTinv}), we
may write the pole denominators in terms of a principle value and a
$\delta$-function contribution:
\begin{equation}
   {1\over u-u_i} \to {\rm P}\,{1\over u-u_i}
   + \eta_i\,\delta(u-u_i) \,.
\end{equation}
Here, $\eta_i$ is a complex number which depends on the
regularization prescription. For instance, one may choose one of the
following regularizations (with $\delta\to +0$)
\begin{eqnarray}
   {u-u_i+\kappa\delta\over (u-u_i)^2 + \delta^2}
   &\quad\to\quad& \eta_i = \kappa\pi \,, \nonumber\\
   {1\over u-u_i\mp i\delta} &\quad\to\quad& \eta_i = \pm i\pi \,.
\end{eqnarray}
One may also choose the principal value prescription itself, in which
case $\eta_i=0$. We write the regularized form of the Borel transform
as
\begin{equation}
   \widetilde F(u) = \widetilde F_{\rm reg}(u)
   + \sum_i \eta_i\,r_i\,\delta(u-u_i) \,,
\end{equation}
where by definition $\widetilde F_{\rm reg}(u)$ contains the pole
terms regularized with a principle value prescription, and all
ambiguity resulting from the freedom to use a different prescription
resides in the $\delta$-function contributions. The inverse Borel
transformation then leads to
\begin{equation}\label{Fstar}
   F(\alpha_s) = F_{\rm reg}(\alpha_s) + \sum_i \eta_i\,r_i\,
   \exp\bigg( -{4\pi u_i\over\beta_0\,\alpha_s(\mu)} \bigg)
   \simeq F_{\rm reg}(\alpha_s) + \sum_i \eta_i\,r_i\,\bigg(
   {\Lambda_{\rm QCD}\over\mu} \bigg)^{2 u_i} \,,
\end{equation}
where $F_{\rm reg}(\alpha_s)$ is the inverse Borel transform of
$\widetilde F_{\rm reg}(u)$. In the last step, we have used the
one-loop expression
\begin{equation}
   \alpha_s(\mu) = {4\pi\over\beta_0\,\ln(\mu^2/\Lambda_{\rm QCD}^2)}
\end{equation}
for the running coupling constant.\footnote{Note that the last
relations in (\ref{Fstar}) and (\ref{DeltaF}) become exact in the
large-$N_f$ limit.}
These definitions make explicit the fact that terms which depend on
the regularization prescription are exponentially small in the
coupling constant, i.e.\ they have the form of power corrections. The
leading asymptotic behaviour is determined by the nearest infrared
renormalon pole at $u=u_1$. We define the renormalon ambiguity
$\Delta F$ as the coefficient of $\eta_1$:
\begin{equation}\label{DeltaF}
   \Delta F = r_1\,\exp\bigg(
   -{4\pi u_1\over\beta_0\,\alpha_s(\mu)} \bigg)
   \simeq r_1\,\bigg( {\Lambda_{\rm QCD}\over\mu} \bigg)^{2 u_1} \,.
\end{equation}
It is a measure of the intrinsic ambiguity in the quantity $F$
arising from the necessity to regularize the divergent behaviour of
perturbation theory in large orders. It is the purpose of this paper
to trace how these leading (in powers of $\Lambda_{\rm QCD}/\mu$)
ambiguities cancel in HQET predictions for physical quantities.

Although the appearance of renormalons can hardly be doubted on
physical grounds, a rigorous proof of their existence does not exist
even in field theories that are much simpler than QCD. For this
reason, various forms of large-$N$ expansions have become the
state-of-the-art approach to study renormalon singularities. In QCD,
one uses $1/N_f$ as an expansion parameter, where $N_f$ is the number
of light quark flavours. In the large-$N_f$ limit, the insertions of
fermion loops in a gluon propagator are the only higher-order
contributions that have to be retained in the perturbative expansion,
since they involve powers of $N_f\,\alpha_s=O(N_f^0)$. Unfortunately,
QCD in the large-$N_f$ limit is not an asymptotically free theory;
the first coefficient of the $\beta$-function becomes negative for
$N_f>33/2$. However, it is believed that although the $1/N_f$
expansion is not adequate to describe the dynamics of QCD, it can
still be used to locate the position of the renormalon poles in the
Borel plane. In other words, the hope is that tracing the fermionic
contribution to the $\beta$-function one gets the remaining
contributions for free, and that using the correct value of $\beta_0$
in (\ref{BTinv}) gives the right result. Although there exists no
proof of this assertion, we will accept it as a working hypothesis.

In the large-$N_f$ limit, the summation of bubbles can be performed
directly on the gluon propagator. In Landau gauge, and after
renormalization of the fermion loops, the Borel transform of the
resummed propagator takes the form \cite{BBren,Bene}
\begin{equation}\label{Dab}
   \widetilde D_{ab}^{\mu\nu}(k,u) = i\delta_{ab}\,
   \Bigg( {e^C\over\mu^2} \Bigg)^{-u}\,
   {k^\mu k^\nu - g^{\mu\nu} k^2\over (-k^2)^{2+u}} \,,
\end{equation}
where $\mu$ is the renormalization scale, and $C$ is a
scheme-dependent constant. In the $\overline{\rm MS}$ scheme,
$C=-5/3$. Consider now an arbitrary correlation function without
external gluons. To order $1/N_f$, all its dependence on the coupling
constant $\alpha_s$ comes from diagrams containing one resummed gluon
propagator. The Borel transform of such diagrams is simply obtained
by using the Borel transformed propagator (\ref{Dab}) instead of the
usual propagator.

\begin{figure}[h]
   \vspace{1.5cm}
   \epsfxsize=5cm
   \centerline{\epsffile{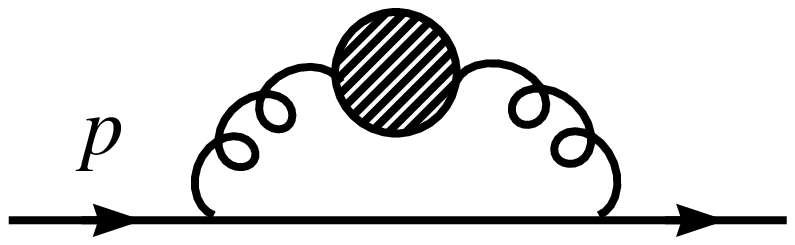}}
   \vspace{-2.5cm}
   \centerline{\parbox{13cm}{\caption{\label{fig:1}
Borel transform of the heavy quark self-energy to order $1/N_f$. The
resummed gluon propagator (\protect\ref{Dab}) is denoted by the
dashed bubble.}}}
\end{figure}

Following Beneke and Braun \cite{BBren}, let us then consider the
structure of infrared renormalons in the pole mass and on-shell
wave-function renormalization of a heavy quark. In terms of the
self-energy $\Sigma(\rlap{\,/}p)$, one has
\begin{equation}
   m_Q^{\rm pole} = m_Q + \Sigma(\rlap{\,/}p)
    \Big|_{\rlap/p=m_Q^{\rm pole}} \,,\qquad
   1 - Z_Q^{-1} = {\partial\Sigma(\rlap{\,/}p)\over
    \partial\rlap{\,/}p} \bigg|_{\rlap/p=m_Q^{\rm pole}} \,,
\end{equation}
where $m_Q$ is the bare mass. In general, these are complicated
implicit equations. However, since the self-energy is of order
$1/N_f$ with respect to the bare mass, one can replace $m_Q^{\rm
pole}$ by $m_Q$ on the right-hand side, thereby neglecting terms of
order $1/N_f^2$. We work in Landau gauge and use dimensional
regularization. By evaluating the diagram depicted in
Fig.~\ref{fig:1},
we obtain for the Borel transform of the self-energy the relations
\begin{eqnarray}
   \widetilde\Sigma(\rlap{\,/}p,u)\Big|_{\rlap/p=m_Q} &=&
    {C_F\,m_Q\over\beta_0}\,(d-1)\,e^{-C u}\,(4\pi)^{2-d/2}\,
    \bigg( {m_Q\over\mu} \bigg)^{d-4-2u} \nonumber\\
   &&\times (d-2-2u)\,{\Gamma(2-\frac{d}{2}+u)\,\Gamma(d-3-2u)
    \over\Gamma(d-1-u)} + O(N_f^{-2}) \,, \nonumber\\
   {\partial\widetilde\Sigma(\rlap{\,/}p,u)\over
    \partial\rlap{\,/}p} \bigg|_{\rlap/p=m_Q} &=&
    - {(1+u)\over m_Q}\,\widetilde\Sigma(\rlap{\,/}p,u)
    \Big|_{\rlap/p=m_Q} + O(N_f^{-2}) \,,
\end{eqnarray}
where $C_F=4/3$, and $d$ denotes the number of space-time dimensions.
For generic $u$, one can evaluate these expressions for $d=4$. The
positions of renormalons are determined by the $\Gamma$-functions in
the numerator. There are infrared renormalons at positive
half-integer values of $u$, as well as ultraviolet renormalons at
negative integer values of $u$. For $u=0$, the self-energy and its
derivative are ultraviolet divergent in $d=4$ dimensions. One can
subtract the ultraviolet divergence by subtracting the pole at $u=0$
after setting $d=4$ \cite{BBren}. This determines the renormalized
Borel transform up to a scheme-dependent function $R(u)$, which is
entire in the Borel plane if a renormalization scheme with analytic
counterterms (such as $\overline{\rm MS}$) is employed. The result is
\begin{eqnarray}\label{mpoleu}
   \widetilde m_Q^{\rm pole}(u) = m_Q^{\rm R}\,\bigg\{ \delta(u)
   &+& {C_F\over\beta_0}\,\bigg[ 6\,e^{-C u}\,\bigg(
    {\mu\over m_Q} \bigg)^{2u} (1-u)\,
    {\Gamma(u)\,\Gamma(1-2u)\over\Gamma(3-u)}
    - {3\over u} + R_m(u) \bigg] \nonumber\\
   &+& O(N_f^{-2}) \bigg\} \,,
\end{eqnarray}
and
\begin{eqnarray}
   \widetilde Z_Q^{\rm R}(u) = \delta(u)
   &+& {C_F\over\beta_0}\,\bigg[ - 6\,e^{-C u}\,
    \bigg( {\mu\over m_Q} \bigg)^{2u}(1-u^2)\,
    {\Gamma(u)\,\Gamma(1-2u)\over\Gamma(3-u)}
    + {3\over u} + R_Z(u) \bigg] \nonumber\\
   &+& O(N_f^{-2}) \,,
\end{eqnarray}
where $m_Q^{\rm R}$ is the renormalized mass. The first expression
has been derived in Ref.~\cite{BBren}. Note that to order $1/N_f$ the
choice of $m_Q$ in the parentheses on the right-hand side is
arbitrary. The functions $R_m(u)$ and $R_Z(u)$ depend on the
renormalization scheme specified by the superscript ``R''. In the
$\overline{\rm MS}$ scheme, one has $R_m(u) = -5/2 + O(u)$ and
$R_Z(u) = 11/2 + O(u)$. For the discussion of renormalon
singularities these functions are irrelevant. The asymptotic
behaviour of the perturbative expansions for $m_Q^{\rm pole}$ and
$Z_Q$ is determined by the nearest infrared renormalon pole, which is
located at $u=1/2$. According to (\ref{DeltaF}), it leads to
intrinsic ambiguities given by
\begin{eqnarray}\label{Dmpole}
   \Delta m_Q^{\rm pole} &=& - {2 C_F\over\beta_0}\,
    e^{-C/2}\,\Lambda_{\rm QCD} + O(N_f^{-2}) \,, \nonumber\\
   \Delta Z_Q &=& {3 C_F\over\beta_0}\,e^{-C/2}\,
    {\Lambda_{\rm QCD}\over m_Q} + O(N_f^{-2}) \,.
\end{eqnarray}
Note that the product $e^{-C/2}\,\Lambda_{\rm QCD}$ is
scheme-independent.

After this lengthy introduction into the problem, let us now turn to
the construction of the HQET \cite{EiHi}--\cite{Mann}. A heavy quark
interacting with light degrees of freedom inside a hadron is almost
on-shell. It is then natural to split its momentum into a ``large''
and a ``small'' piece according to $p_Q = m_Q\,v + k$, where $v$ is
the velocity of the hadron, and $m_Q$ is some choice of the heavy
quark mass discussed in detail below. For the moment let us just
require that the components of the residual momentum $k$ are much
smaller than $m_Q$. One then proceeds by introducing a
velocity-dependent heavy quark field $h_v(x)$, which is related to
the original field $Q(x)$ by
\begin{equation}\label{redef}
   h_v(x) = \exp(i m_Q\,v\cdot x)\,{(1+\rlap/v)\over 2}\,Q(x) \,.
\end{equation}
The effective Lagrangian for $h_v$ reads \cite{Geor,Mann,FNL}
\begin{equation}\label{Leff}
   {\cal L}_{\rm eff} = \bar h_v\,(i v\cdot D-\delta m)\,h_v
   + \ldots \,,
\end{equation}
where $\delta m$ is the residual mass term for the heavy quark in the
effective theory. It appears since there is a freedom in the choice
of the expansion parameter $m_Q$ in (\ref{redef}). One can show that
in physical matrix elements only the combination $(m_Q+\delta m)$
appears, i.e.\ different choices of $m_Q$ are compensated by
different values of $\delta m$ \cite{FNL}. The ellipses in
(\ref{Leff}) represent terms that contain additional powers of $i
D^\mu/m_Q$ or $\delta m/m_Q$. If one arranges things in such a way
that the components of $k$ and $\delta m$ are of order $\Lambda_{\rm
QCD}$ and independent of $m_Q$, this construction provides a
systematic expansion in powers of $\Lambda_{\rm QCD}/m_Q$. Moreover,
the leading terms in the effective Lagrangian (\ref{Leff}) are then
invariant under a spin-flavour symmetry group. To this end, the heavy
quark mass $m_Q$ used in the field redefinition (\ref{redef}) must be
a ``physical'' mass such as the pole mass, the mass of the lightest
hadron that contains the heavy quark, or any other definition that
differs from the pole mass by an amount of order $\Lambda_{\rm QCD}$.
The residual mass term is given by $\delta m = m_Q^{\rm pole} - m_Q$;
i.e.\ if one chooses the pole mass to construct the HQET, the
residual mass vanishes, and to any finite order in perturbation
theory the effective heavy quark propagator has a pole at $k=0$.
However, from our previous considerations we know that there is an
intrinsic ambiguity of order $\Lambda_{\rm QCD}$ in the definition of
the pole mass, once non-perturbative effects are taken into account.
Hence, if one wants to write down the Lagrangian of the HQET without
specifying a particular Borel summation prescription, one can do this
for the price of an ambiguous residual mass term \cite{BBren}. To be
specific, let us construct the HQET using the heavy quark mass
defined with a principle value prescription to regularize the poles
in the Borel plane [cf.~(\ref{Fstar})]. It then follows that
\begin{equation}\label{dmexpr}
   \delta m = \eta_1\,\Delta m_Q^{\rm pole} = - \eta_1\,
   {2 C_F\over\beta_0}\,e^{-C/2}\,\Lambda_{\rm QCD}
   + O(N_f^{-2}) \,.
\end{equation}
The ambiguity associated with the definition of the pole mass shows
up in the form of an ambiguous parameter in the effective Lagrangian
(\ref{Leff}). At first sight this may seem a problem: How can one
derive unambiguous predictions from a Lagrangian that contains an
ambiguous parameter? The answer is that the effective theory has to
be matched onto the full theory at some large momentum scale. In this
process there appear coefficient functions multiplying the operators
of the HQET. The ambiguous residual mass term is required to cancel
ambiguities in these coefficient functions. The important point to
note is that $\delta m$ is independent of $m_Q$ and thus does not
break the flavour symmetry of the effective Lagrangian. The way in
which the residual mass enters the $1/m_Q$ expansion has been
investigated in Ref.~\cite{FNL}.

Let us come back, at this point, to the original formulation of
Wilson's OPE \cite{Wils}, in which renormalons never appear.
Introducing a hard factorization scale $\mu$ in the construction of
the HQET would yield a residual mass term of the form $\delta
m\sim\mu\,\alpha_s(\mu)$. Likewise, hadronic matrix elements in the
effective theory as well as the Wilson coefficient functions would
have a power-like dependence on $\mu$, in such a way that the
factorization scale disappears from the final predictions for
physical quantities. This is the content of the renormalization-group
equation. In this formulation, the parameters of the theory are not
plagued by ambiguities, but they depend on the arbitrary parameter
$\mu$. Moreover, the precise form of this dependence (for instance,
the coefficient of the $\mu\,\alpha_s(\mu)$ term in $\delta m$)
depends on how exactly the hard cutoff is implemented in Feynman
diagrams. Hence, there is a similar arbitrariness in the definition
of these parameters as in the case of the practical form of the OPE,
which contains renormalons. Finally, we note that chosing $\mu=m_Q$
as a factorization scale (as it was proposed in
Refs.~\cite{Bigiren,Patr}) breaks the flavour symmetry of the
effective Lagrangian (through the residual mass term) and is thus not
a viable choice in processes that involve more than one heavy quark
flavour.

\section{Heavy-to-Heavy Transition Matrix Elements}
\label{sec:3}

In this section we investigate how renormalons appear in the hadronic
matrix elements that describe current-induced transitions between two
hadrons containing heavy quarks with masses $m_1$ and $m_2$. These
matrix elements are of the form $\langle H_2(v_2)|\,\bar Q_2\,
\Gamma\, Q_1\,|H_1(v_1)\rangle$, where $v_1$ and $v_2$ denote the
velocities of the hadrons. The quantum numbers of the light degrees
of freedom are assumed to be the same in the initial and final state,
but are otherwise arbitrary. We will restrict ourselves to the cases
of vector or axial vector currents ($\Gamma=\gamma^\mu$ or
$\gamma^\mu\gamma_5$). This covers semileptonic weak decays such as
$\bar B\to D^{(*)}\ell\,\bar\nu$ and $\Lambda_b\to\Lambda_c\,\ell\,
\bar\nu$.

In the HQET, the currents which mediate these transitions obey an
expansion in a series of local operators multiplied by coefficient
functions. These functions depend upon the heavy quark masses, the
renormalization scale, and the hadron velocity product $w=v_1\cdot
v_2$. A particular property of heavy-to-heavy transitions is that the
coefficients of the operators of dimension four are all related to
the coefficients of the dimension-three operators \cite{Repa}. The
reason for this is an invariance of the effective theory under
reparametrization of the heavy quark momentum \cite{LuMa}. As an
example, we give the exact form of the expansion of the vector
current to order $1/m_Q$ \cite{Repa}:
\begin{eqnarray}\label{current}
   \bar Q_2\,\gamma^\mu Q_1 &\to&  C_1^V\,\Bigg\{
    \bar h_{v_2}\gamma^\mu h_{v_1}
    + {1\over 2 m_1}\,\bar h_{v_2}\gamma^\mu
     i\rlap{\,/}{\cal D}_1\,h_{v_1}
    - {1\over 2 m_2}\,\bar h_{v_2}\,
     i\overleftarrow{\rlap{\,/}{\cal D}_2}\,\gamma^\mu h_{v_1}
     \Bigg\} \nonumber\\
   &+& {\partial C_1^V\over\partial w}\,\Bigg\{
    {1\over m_1}\,\bar h_{v_2}\gamma^\mu
     i v_2\!\cdot\!{\cal D}_1\,h_{v_1}
    - {1\over m_2}\,\bar h_{v_2}\,
     i v_1\!\cdot\!\overleftarrow{{\cal D}_2}\,
     \gamma^\mu h_{v_1} \Bigg\} \nonumber\\
   &+& C_2^V\,\Bigg\{
    \bar h_{v_2}\,v_1^\mu\,h_{v_1}
    + {1\over 2 m_1}\,\bar h_{v_2}\,v_1^\mu\,
     i\rlap{\,/}{\cal D}_1\,h_{v_1}
    - {1\over 2 m_2}\,\bar h_{v_2}\,
     i\overleftarrow{\rlap{\,/}{\cal D}_2}\,v_1^\mu\,h_{v_1}
    + {1\over m_1}\,\bar h_{v_2}\,i{\cal D}_1^\mu\,h_{v_1}
     \Bigg\} \nonumber\\
   &+& {\partial C_2^V\over\partial w}\,\Bigg\{
    {1\over m_1}\,\bar h_{v_2}\,v_1^\mu\,
     i v_2\!\cdot\!{\cal D}_1\,h_{v_1}
    - {1\over m_2}\,\bar h_{v_2}\,
     i v_1\!\cdot\!\overleftarrow{{\cal D}_2}\,v_1^\mu\,h_{v_1}
     \Bigg\} \nonumber\\
   &+& C_3^V\,\Bigg\{
    \bar h_{v_2}\,v_2^\mu\,h_{v_1}
    + {1\over 2 m_1}\,\bar h_{v_2}\,v_2^\mu\,
     i\rlap{\,/}{\cal D}_1\,h_{v_1}
    - {1\over 2 m_2}\,\bar h_{v_2}\,
     i\overleftarrow{\rlap{\,/}{\cal D}_2}\,v_2^\mu\,h_{v_1}
    - {1\over m_2}\,\bar h_{v_2}\,i\overleftarrow{{\cal D}_2^\mu}
     \,h_{v_1} \Bigg\} \nonumber\\
   &+& {\partial C_3^V\over\partial w}\,\Bigg\{
    {1\over m_1}\,\bar h_{v_2}\,v_2^\mu\,
     i v_2\!\cdot\!{\cal D}_1\,h_{v_1}
    - {1\over m_2}\,\bar h_{v_2}\,
     i v_1\!\cdot\!\overleftarrow{{\cal D}_2}\,v_2^\mu\,h_{v_1}
     \Bigg\} \nonumber\\
   &+& O\bigg( {1\over m_1^2}, {1\over m_2^2},
    {1\over m_1 m_2} \bigg) \,,
\end{eqnarray}
where $C_i^V=C_i^V(m_1/\mu,m_2/\mu,w)$. A similar expansion with
coefficients $C_i^A$, and with $\gamma_5$ inserted after whatever
object carries the Lorentz index $\mu$, holds for the axial vector
current. The symbols ${\cal D}_i$ represent combinations of a
gauge-covariant derivative and the residual mass term. They are
defined as \cite{FNL}
\begin{equation}
   i{\cal D}_1^\mu = i D^\mu - \delta m\,v_1^\mu \,,\qquad
   i\overleftarrow{{\cal D}_2^\mu} = i\overleftarrow{D^\mu}
   + \delta m\,v_2^\mu \,.
\end{equation}

We will show below that the coefficient functions $C_i^{V,A}$ of the
dimension-three operators contain infrared renormalons at $u=1/2$,
corresponding to power behaviour of order $1/m_1$ or $1/m_2$. In
order for the physical heavy-to-heavy transition amplitudes to be
unambiguous, we have to require that these renormalons be compensated
by ultraviolet renormalons in HQET matrix elements of dimension-four
operators. This requirement is analogous to the renormalization-group
equation in Wilson's OPE. The complete set of dimension-four
operators consists of the local current operators in (\ref{current})
as well as operators containing the time-ordered product of a
dimension-three operator with a $1/m_Q$ insertion from the effective
Lagrangian \cite{Luke}, which at this order is given by
\cite{EiHi,Mann}
\begin{equation}\label{Lfull}
   {\cal L}_{\rm eff} = \bar h_v\,i v\cdot{\cal D}\,h_v
   + {1\over 2 m_Q}\,\bar h_v\,(i{\cal D})^2 h_v
   + C_{\rm mag}(m_Q/\mu)\,{g_s\over 4 m_Q}\,
   \bar h_v\,\sigma_{\mu\nu} G^{\mu\nu} h_v + \ldots \,.
\end{equation}
However, a cancellation of renormalon ambiguities can only occur
between terms that have the structure of matrix elements of local
dimension-three operators. In other words, only the matrix elements
of dimension-four operators that can mix with lower dimensional
operators can contain ultraviolet renormalons.

In heavy-to-heavy transitions, the ultraviolet renormalons in the
matrix elements of the local dimension-four operators can be related
to the infrared renormalon in the pole mass. Using the equation of
motion of the HQET, $i v_1\!\cdot\!{\cal D}_1\,h_{v_1}=0$, as well as
an integration by parts, one can show that \cite{Luke,FNL}
\begin{eqnarray}\label{Lamdm}
   \langle H_2(v_2)|\,\bar h_{v_2}\,\Gamma\,i{\cal D}_1^\alpha\,
   h_{v_1}\,|H_1(v_1)\rangle &=& {\bar\Lambda\over w+1}\,
    \langle H_2(v_2)|\,\bar h_{v_2}\,\Gamma\,
    (w\,v_1^\alpha - v_2^\alpha)\,h_{v_1}\,|H_1(v_1)\rangle
    + \ldots \,, \nonumber\\
   -\langle H_2(v_2)|\,\bar h_{v_2}
   i\overleftarrow{{\cal D}_2^\alpha}\,\Gamma\,h_{v_1}
   |H_1(v_1)\rangle &=& {\bar\Lambda\over w+1}
    \langle H_2(v_2)|\,\bar h_{v_2}(w\,v_2^\alpha - v_1^\alpha)\,
    \Gamma\,h_{v_1} |H_1(v_1)\rangle + \ldots \,, \nonumber\\
\end{eqnarray}
where $\Gamma$ may be an arbitrary Dirac matrix. The ellipses
represent terms that vanish upon contraction with $v_{1\alpha}$ or
$v_{2\alpha}$. These terms cannot be written in the form of matrix
elements of local operators. Hence, as explained above, they must be
free of renormalons. As an example, consider the case of the
ground-state pseudoscalar and vector mesons. There, the matrix
elements of local dimension-three operators can be parametrized in
terms of the Isgur-Wise function \cite{Isgu}:
\begin{equation}\label{IWdef}
   \langle M_2(v_2)|\,\bar h_{v_2} \Gamma\,h_{v_1}\,
   |M_1(v_1)\rangle = - \xi(w,\mu)\,{\rm Tr}\Big\{
   \overline{\cal M}_2(v_2)\,\Gamma\,{\cal M}_1(v_1) \Big\} \,,
\end{equation}
where ${\cal M}(v)$ are the tensor wave functions defined in
Ref.~\cite{Falk}. The matrix elements of local dimension-four
operators can be written as \cite{Luke,FNL}
\begin{eqnarray}
   \langle M_2(v_2)|\,\bar h_{v_2} \Gamma\,i{\cal D}_1^\alpha
   h_{v_1}\,|M_1(v_1)\rangle
   &=& - {\bar\Lambda\over w+1}\,\xi(w,\mu)\,
    {\rm Tr}\Big\{ \overline{\cal M}_2(v_2)\,
    (w\,v_1^\alpha - v_2^\alpha)\,\Gamma\,{\cal M}_1(v_1) \Big\}
    \nonumber\\
   &&\mbox{}+ \xi_3(w,\mu)\,{\rm Tr}\bigg\{ \bigg( \gamma^\alpha
    - {v_1^\alpha + v_2^\alpha\over w+1} \bigg)\,
    \overline{\cal M}_2(v_2)\,\Gamma\,{\cal M}_1(v_1) \bigg\} \,.
    \nonumber\\
\end{eqnarray}
Note that the Feynman rules of the HQET imply that there cannot
appear Dirac matrices next to $\Gamma$ under the trace with the meson
wave functions. Obviously, the structure of the trace associated with
the function $\xi_3(w,\mu)$ is different from the structure of the
trace in (\ref{IWdef}). It follows that $\xi_3(w,\mu)$ does not
contain an ultraviolet renormalon at $u=1/2$. Let us now come back to
the terms shown explicitly in (\ref{Lamdm}). They have the structure
of matrix elements of the local dimension-three operators. For
instance, in the case of the second operator on the right-hand side
in (\ref{current}) one has $\Gamma=\gamma^\mu \gamma_\alpha$, and
between the heavy quark spinors one can replace $(w\,v_1^\alpha -
v_2^\alpha)\,\gamma^\mu\gamma_\alpha$ by $(w+1)\gamma^\mu - 2
v_2^\mu$. The parameter
\begin{equation}\label{Lamdef}
   \bar\Lambda = m_{H_i} - m_i - \delta m
   = m_{H_i} - m_i^{\rm pole} \,;\qquad i=1,2
\end{equation}
denotes the asymptotic value of the difference between the hadron and
heavy quark pole masses, which is flavour-independent. Note that this
parameter is independent of the choice of the expansion parameter
$m_Q$ used in the construction of the HQET \cite{FNL}. However,
because of its dependence on the pole mass it does contain an
ultraviolet renormalon \cite{BBren}. Using (\ref{dmexpr}), we find
that the corresponding ambiguity in $\bar\Lambda$ is given by
\begin{equation}\label{DLambar}
   \Delta\bar\Lambda = -\Delta m_Q^{\rm pole}
   = {2 C_F\over\beta_0}\,e^{-C/2}\,\Lambda_{\rm QCD}
   + O(N_f^{-2}) \,.
\end{equation}

Next consider the matrix elements of the operators containing the
time-ordered product of two local operators. An insertion of the
kinetic operator $(1/2 m_Q)\,\bar h_v\,(i{\cal D})^2 h_v$ into a
matrix element of a local dimension-three operator does not affect
the transformation properties under the Lorentz group and heavy quark
spin symmetry. The effect of such an insertion is simply a
multiplicative renormalization of the original matrix element. We
define a function $K(w,\mu)$ by\footnote{For meson decays, the form
factor $K(w,\mu)$ is usually written as
$K(w,\mu)=2\chi_1(w,\mu)/\xi(w,\mu)$ \cite{Luke}, where $\xi(w,\mu)$
is the Isgur-Wise function \cite{Isgu}.}
\begin{eqnarray}\label{Kdef}
   && \langle H_2(v_2)|\,i\int{\rm d}^4 x\,{\rm T} \Big\{
    \bar h_{v_1}(x)\,(i{\cal D}_1)^2 h_{v_1}(x),
    \bar h_{v_2}(0)\,\Gamma\,h_{v_1}(0) \Big\}\,|H_1(v_1)\rangle
    \nonumber\\
   &=& \langle H_2(v_2)|\,i\int{\rm d}^4 x\,{\rm T} \Big\{
    \bar h_{v_2}(x)\,(i{\cal D}_2)^2 h_{v_2}(x),
    \bar h_{v_2}(0)\,\Gamma\,h_{v_1}(0) \Big\}\,|H_1(v_1)\rangle
    \nonumber\\
   \phantom{ \int } &=& K(w,\mu)\,\langle H_2(v_2)|\,
    \bar h_{v_2}(0)\,\Gamma\,h_{v_1}(0)\,|H_1(v_1)\rangle \,.
\end{eqnarray}
Clearly, these time-ordered products can mix with the dimension-three
operators under renormalization. This is obvious if a dimensionful
regulator is employed. But even in dimensional regularization, it can
be seen from the renormalization-group equation for the function
$K(w,\mu)$, which contains an inhomogeneous term proportional to
$\bar\Lambda$ \cite{MN1}:
\begin{equation}\label{KRGE}
   \mu\,{{\rm d}\over{\rm d}\mu}\,K(w,\mu)
   = - 2\bar\Lambda\,(w-1)\,{\partial\over\partial w}\,
   \gamma_{\rm hh}(w) \,.
\end{equation}
Here $\gamma_{\rm hh}(w)$ denotes the velocity-dependent anomalous
dimension of the currents in the effective theory \cite{Falk}:
\begin{eqnarray}\label{gamhh}
   \gamma_{\rm hh}(w) &=& {C_F\,\alpha_s\over\pi}\,
    \Big[ w\,r(w)-1 \Big] + O(\alpha_s^2) \,; \nonumber\\
   r(w) &=& {1\over\sqrt{w^2-1}}\,\ln\big( w + \sqrt{w^2-1} \big) \,.
\end{eqnarray}
{}From (\ref{KRGE}), it follows that the function $K(w,\mu)$ contains
an ultraviolet renormalon at $u=1/2$. Let us denote its renormalon
ambiguity by $\Delta K(w)$. Vector current conservation implies that
$K(w,\mu)$ must vanish at zero recoil ($w=1$) \cite{Luke}, and this
requires that
\begin{equation}\label{DK1}
   \Delta K(1) = 0 \,.
\end{equation}
Finally, we note that insertions of the chromo-magnetic operator
change the transformation properties of matrix elements in such a way
that there is no mixing with matrix elements of lower dimensional
operators \cite{Luke}. Hence, the HQET functions that parameterize
these matrix elements are free of ultraviolet renormalons. This is
again special to the case of heavy-to-heavy transitions, where the
spin symmetry applies to both the initial and final state.

We can now equate the infrared and ultraviolet renormalon ambiguities
to derive a set of conditions that have to be fulfilled in order to
obtain unambiguous predictions for the physical heavy-to-heavy
transition form factors. Separating the terms associated with
different Lorentz structures, we obtain from (\ref{current}),
(\ref{Lamdm}), (\ref{DLambar}) and (\ref{Kdef}):
\begin{eqnarray}
   \Delta C_1 &=& {\Delta m_{\rm pole}\over 2}\,\bigg(
    {1\over m_1} + {1\over m_2} \bigg) \bigg\{ {w\pm 1\over w+1}\,
    C_1 + 2(w-1)\,{\partial C_1\over\partial w} \bigg\}
    - {\Delta K\over 2}\,\bigg( {1\over m_1} + {1\over m_2}
    \bigg)\,C_1 \,, \nonumber\\
   \Delta C_2 &=& {\Delta m_{\rm pole}\over 2}\,\bigg(
    {1\over m_1} + {1\over m_2} \bigg) \bigg\{ {w\pm 1\over w+1}\,
    C_2 + 2(w-1)\,{\partial C_2\over\partial w} \bigg\}
    + {\Delta m_{\rm pole}\over m_1}\,{w\mp 1\over w+1}\,C_2
    \nonumber\\
   &&\mbox{}- {\Delta m_{\rm pole}\over m_2}\,{1\over w+1}\,
    \big( C_1 \pm C_2 + C_3 \big)
    - {\Delta K\over 2}\,\bigg( {1\over m_1} + {1\over m_2}
    \bigg)\,C_2 \,, \nonumber\\
   \Delta C_3 &=& {\Delta m_{\rm pole}\over 2}\,\bigg(
    {1\over m_1} + {1\over m_2} \bigg) \bigg\{ {w\pm 1\over w+1}\,
    C_3 + 2(w-1)\,{\partial C_3\over\partial w} \bigg\}
    + {\Delta m_{\rm pole}\over m_2}\,{w\mp 1\over w+1}\,C_3
    \nonumber\\
   &&\mbox{}\mp {\Delta m_{\rm pole}\over m_1}\,{1\over w+1}\,
    \big( C_1 \pm C_2 + C_3 \big)
    - {\Delta K\over 2}\,\bigg( {1\over m_1} + {1\over m_2}
    \bigg)\,C_3 \,.
\end{eqnarray}
We use a short-hand notation where we omit the superscript $V$ or $A$
on the coefficient functions, and where upper (lower) signs refers to
the coefficients of the vector (axial vector) current. In total,
there are thus six relations. We find it useful to solve for $\Delta
K$ using the relation for $\Delta C_1^A$, and to eliminate $\Delta K$
from the remaining relations. This leads to
\begin{equation}\label{DelK}
   \Delta K = \Delta m_{\rm pole}\,(w-1)\,\bigg\{ {1\over w+1}
   + 2\,{\partial\over\partial w}\,\ln C_1^A \bigg\}
   - 2\,\bigg( {1\over m_1} + {1\over m_2} \bigg)^{-1}\,
   {\Delta C_1^A\over C_1^A} \,,
\end{equation}
as well as
\begin{eqnarray}\label{conrel}
   {1\over\Delta m_{\rm pole}}\,\bigg( {\Delta C_1^V\over C_1^V}
   - {\Delta C_1^A\over C_1^A} \bigg)
   &=& \bigg( {1\over m_1} + {1\over m_2} \bigg)\bigg\{
    {1\over w+1}+ (w-1)\,{\partial\over\partial w}\,
    \ln{C_1^V\over C_1^A} \bigg\} \,, \nonumber\\
   {1\over\Delta m_{\rm pole}}\,\bigg( \Delta C_2
   - {C_2\over C_1}\,\Delta C_1 \bigg) &=& -{1\over m_2}\,
    {1\over w+1}\,\big( C_1 \pm C_2 + C_3 \big)
    + {1\over m_1}\,{w\mp 1\over w+1}\,C_2 \nonumber\\
   &&\mbox{}+ \bigg( {1\over m_1} + {1\over m_2} \bigg)\,(w-1)\,
    C_2\,{\partial\over\partial w}\,\ln{C_2\over C_1} \,,
    \nonumber\\
   {1\over\Delta m_{\rm pole}}\,\bigg( \Delta C_3
   - {C_3\over C_1}\,\Delta C_1 \bigg) &=& \mp {1\over m_1}\,
    {1\over w+1}\,\big( C_1 \pm C_2 + C_3 \big)
    + {1\over m_2}\,{w\mp 1\over w+1}\,C_3 \nonumber\\
   &&\mbox{}+ \bigg( {1\over m_1} + {1\over m_2} \bigg)\,(w-1)\,
    C_3\,{\partial\over\partial w}\,\ln{C_3\over C_1} \,.
\end{eqnarray}
Equation (\ref{DelK}) determines the structure of the ultraviolet
renormalon in the hadronic form factor $K(w,\mu)$ in terms of the
infrared renormalon in the pole mass and in the coefficient function
$C_1^A$. Since we are not able to calculate the hadronic form factor
$K(w,\mu)$ from first principles (not even using a $1/N_f$
expansion), we cannot check this relation, but we can use it to
compute $\Delta K$ given a calculation of $\Delta m_{\rm pole}$ and
$\Delta C_1^A$. The remaining five relations in (\ref{conrel}) form a
set of consistency conditions involving only the infrared renormalons
in the pole mass and in the coefficient functions. Unless these
conditions are satisfied, a compensation of infrared and ultraviolet
renormalons is not possible. The existence of such relations is
non-trivial and is a consequence of the strong constraints imposed by
heavy quark symmetry on the structure of the weak decay form factors
in heavy-to-heavy transitions.

\begin{figure}[h]
   \vspace{1.5cm}
   \epsfxsize=11cm
   \centerline{\epsffile{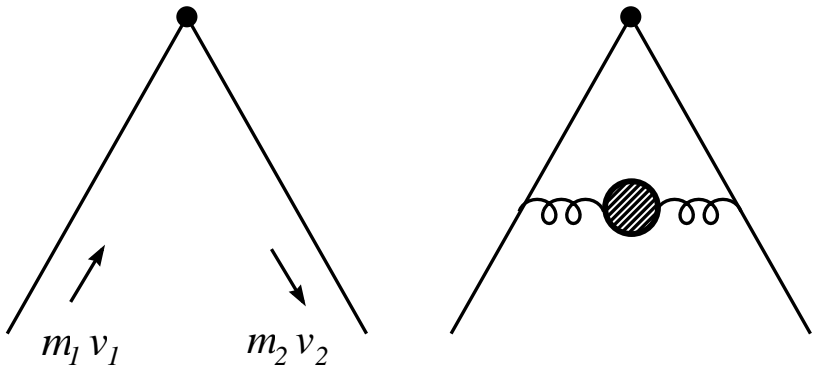}}
   \centerline{\parbox{13cm}{\caption{\label{fig:2}
Vertex contribution to the matching calculation of the coefficient
functions of heavy-heavy currents.}}}
\end{figure}

The above results are exact to all orders in $1/N_f$. In the
large-$N_f$ limit, they simplify since the renormalon ambiguities are
of order $1/N_f$, and we can use the fact that $C_i = \delta_{i1} +
O(1/N_f)$. This leads to
\begin{equation}\label{DeltaK}
   \Delta K = {w-1\over w+1}\,\Delta m_{\rm pole}
   - 2\,\bigg( {1\over m_1} + {1\over m_2} \bigg)^{-1}\,
   \Delta C_1^A + O(N_f^{-2}) \,,
\end{equation}
and
\begin{eqnarray}\label{consi}
   \Delta C_1^V - \Delta C_1^A &=& {\Delta m_{\rm pole}\over w+1}\,
    \bigg( {1\over m_1} + {1\over m_2} \bigg) + O(N_f^{-2}) \,,
    \nonumber\\
   \Delta C_2^{V,A} &=& - {1\over w+1}\,
    {\Delta m_{\rm pole}\over m_2} + O(N_f^{-2}) \,, \nonumber\\
   \Delta C_3^{V,A} &=& \mp {1\over w+1}\,
    {\Delta m_{\rm pole}\over m_1} + O(N_f^{-2}) \,.
\end{eqnarray}
Let us now check these relations with an explicit calculation of the
asymptotic behaviour of the coefficient functions to order $1/N_f$.
The coefficients are obtained by comparing matrix elements in the
HQET with matrix elements in the full theory at some reference scale
$\mu$. This matching procedure is independent of the external states,
and it is most economic to evaluate the matrix elements with on-shell
quark states. If one uses dimensional regularization, all loop
diagrams in the HQET vanish, i.e.\ the coefficient functions are
simply given by the on-shell vertex functions of the full theory
\cite{EiHi,review}. Hence, the Borel-transformed coefficient
functions are obtained by evaluating the diagram shown in
Fig.~\ref{fig:2} supplemented by wave-function renormalization.
Setting $d=4$ in the final result, we obtain
\begin{eqnarray}\label{Cres}
   \widetilde C_1^{V,A}(u) &=& \delta(u) - {3 C_F\over\beta_0}\,
    e^{-C u}\,\bigg[ \bigg( {\mu\over m_1} \bigg)^{2u}
    + \bigg( {\mu\over m_2} \bigg)^{2u} \bigg]\,
    (1-u^2)\,{\Gamma(u)\,\Gamma(1-2u)\over\Gamma(3-u)} \nonumber\\
   &&\mbox{}+ {C_F\over\beta_0}\,
    e^{-C u}\,\bigg( {\mu^2\over m_1 m_2} \bigg)^u\,
    {\Gamma(u)\,\Gamma(1-2u)\over\Gamma(2-u)}\,\Bigg\{
    2 \Big[ (1+u)\,w\pm u \Big]\,F_{11}^{1+u} \nonumber\\
   &&\qquad\qquad\mbox{} + u \bigg( {m_1\over m_2}\,F_{21}^{1+u}
    + {m_2\over m_1}\,F_{12}^{1+u} \bigg)
    + 2 {(1-u)(1-2u)\over (2-u)}\,F_{11}^u \Bigg\} \nonumber\\
   &&\mbox{}+ {C_F\over\beta_0}\,\bigg\{ - {2\over u}\,
    \Big[ w\,r(w)-1 \Big] + R_{V,A}(u) \bigg\} + O(N_f^{-2}) \,,
    \nonumber\\
   \widetilde C_2^{V,A}(u) &=& - {2 C_F\over\beta_0}\,
    e^{-C u}\,\bigg( {\mu^2\over m_1 m_2} \bigg)^u\,
    {\Gamma(1+u)\,\Gamma(1-2u)\over\Gamma(2-u)} \nonumber\\
   &&\times\Bigg\{ F_{12}^{1+u} - {(1-2u)\over 3(2-u)}\,
    \bigg( F_{22}^{1+u} \pm {2 m_1\over m_2}\,F_{31}^{1+u}
    \bigg) \Bigg\} + O(N_f^{-2}) \,, \nonumber\\
   \widetilde C_3^{V,A}(u) &=& \mp {2 C_F\over\beta_0}\,
    e^{-C u}\,\bigg( {\mu^2\over m_1 m_2} \bigg)^u\,
    {\Gamma(1+u)\,\Gamma(1-2u)\over\Gamma(2-u)} \nonumber\\
   &&\times\Bigg\{ F_{21}^{1+u} - {(1-2u)\over 3(2-u)}\,
    \bigg( F_{22}^{1+u} \pm {2 m_2\over m_1}\,F_{13}^{1+u}
    \bigg) \Bigg\} + O(N_f^{-2}) \,,
\end{eqnarray}
where
\begin{equation}
   F_{ab}^c = {\Gamma(a+b)\over\Gamma(a)\,\Gamma(b)}\,
   \int\limits_0^1\!{\rm d}x\,x^{a-1} (1-x)^{b-1}\,
   \bigg[ x^2\,{m_1\over m_2} + (1-x)^2\,{m_2\over m_1}
   + 2 x (1-x)\,w \bigg]^{-c} \,.
\end{equation}
The terms in the last row in $\widetilde C_1^{V,A}(u)$ come from a
renormalization of the ultraviolet divergences for $u=0$. The
coefficient of the $1/u$ pole is proportional to the one-loop
coefficient of the velocity-dependent anomalous dimension in
(\ref{gamhh}). The detailed form of the entire function $R_{V,A}(u)$
is irrelevant for our discussion. We note that a check of the
complicated expressions (\ref{Cres}) is provided by an expansion
around $u=0$, from which we recover the one-loop results for the
coefficient functions derived in Ref.~\cite{QCD}.

It is a simple exercise to extract the residues of the renormalon
poles at $u=1/2$ from the above expressions. The relevant parameter
integrals are given by
\begin{equation}
   F_{12}^{3/2} = {2\over w+1}\,\sqrt{m_1\over m_2} \,,\qquad
   F_{21}^{3/2} = {2\over w+1}\,\sqrt{m_2\over m_1} \,,\qquad
   F_{11}^{3/2} = {1\over 2}\,\big( F_{12}^{3/2} +
   F_{21}^{3/2} \big) \,.
\end{equation}
Using (\ref{DeltaF}) and (\ref{Dmpole}), we then compute the
corresponding renormalon ambiguities in the coefficient functions. We
find that the consistency conditions (\ref{consi}) are indeed
satisfied. In particular, we note that
\begin{eqnarray}\label{C1VC1A}
   \Delta C_1^V &=& {\Delta m_{\rm pole}\over 4}\,{3w+1\over w+1}\,
    \bigg( {1\over m_1} + {1\over m_2} \bigg)
    + O(N_f^{-2}) \,, \nonumber\\
   \Delta C_1^A &=& {\Delta m_{\rm pole}\over 4}\,{3(w-1)\over w+1}
    \,\bigg( {1\over m_1} + {1\over m_2} \bigg)
    + O(N_f^{-2}) \,,
\end{eqnarray}
so that the difference obeys the first relation given in
(\ref{consi}). Moreover, for the ultraviolet renormalon ambiguity in
the function $K(w,\mu)$ we obtain from (\ref{DeltaK})
\begin{equation}
   \Delta K(w) = -{w-1\over w+1}\,{\Delta m_{\rm pole}\over 2}
   + O(N_f^{-2}) \,.
\end{equation}
Note that $\Delta K(w)$ vanishes for $w=1$, as required by vector
current conservation [see (\ref{DK1})].

We have emphasized above that the appearance of renormalons does not
obscure the structure of the heavy quark expansion. In particular,
Luke's theorem \cite{Luke}, which concerns the vanishing of
first-order power corrections at zero recoil, remains unaffected. Let
us illustrate this important fact with the example of the meson decay
form factor $h_{A_1}(w)$ defined by \cite{review}
\begin{equation}
   \langle D^*(v_2,\epsilon)|\,\bar c\,\gamma^\mu\gamma_5 b\,
   |\bar B(v_1)\rangle = \sqrt{m_B m_{D^*}}\,(w+1)\,h_{A_1}(w)\,
   \epsilon^{*\mu} + \ldots \,,
\end{equation}
where $w=v_1\cdot v_2$, and $\epsilon$ denotes the polarization
vector of the $D^*$-meson. This form factor plays a crucial role in
the extraction of $|\,V_{cb}|$ from the extrapolation of the $\bar
B\to D^*\ell\,\bar\nu$ decay rate to zero recoil. One obtains
\cite{Vcb}
\begin{equation}
   \lim_{w\to 1} {1\over\sqrt{w^2-1}}\,
   {{\rm d}\Gamma(\bar B\to D^*\ell\,\bar\nu)\over{\rm d}w}
   = {G_F^2\over 4\pi^3}\,(m_B-m_{D^*})^2\,m_{D^*}^3\,|\,V_{cb}|^2\,
   |h_{A_1}(1)|^2 \,.
\end{equation}
The important point is that $h_{A_1}(1)$ is protected by Luke's
theorem against first-order power corrections \cite{Luke}. It follows
that
\begin{equation}\label{hA1}
   h_{A_1}(1) = \eta_A  + O(1/m_Q^2) \,;\qquad
   \eta_A = C_1^A(w=1) \,,
\end{equation}
where we use $m_Q$ as a generic notation for $m_c$ or $m_b$. The
presence of an infrared renormalon at $u=1/2$ in the short-distance
coefficient $\eta_A$ would spoil this non-renormalization theorem.
However, from (\ref{DeltaK}) and (\ref{DK1}) it follows that the
infrared renormalon at $u=1/2$ in $C_1^A$ vanishes at zero recoil.
Our explicit result (\ref{C1VC1A}) confirms this to order $1/N_f$.
Thus, the theoretical uncertainty in the determination of
$|\,V_{cb}|$ is of order $1/m_Q^2$; it is not affected by a
renormalon ambiguity of order $1/m_Q$. Note, however, that the
expression for $\widetilde C_1^A(u)$ in (\ref{Cres}) contains a
renormalon pole at $u=1$, which does not vanish at zero recoil. The
corresponding ambiguity in $\eta_A$ is given by
\begin{equation}\label{DetaA}
   \Delta\eta_A = {C_F\over 2\beta_0}\,e^{-C}\,\Lambda_{\rm QCD}^2
   \,\bigg( {1\over m_1} + {1\over m_2} \bigg)^2 + O(N_f^{-2}) \,.
\end{equation}
This infrared renormalon must be compensated by an ultraviolet
renormalon in the terms of order $1/m_Q^2$ in (\ref{hA1}). For
completeness, let us also study the renormalization of the vector
current at zero recoil. There, the relevant combination of
coefficient functions is
\begin{equation}
   \eta_V = C_1^V(w=1) + C_2^V(w=1) + C_3^V(w=1) \,.
\end{equation}
We find that the leading renormalon pole in the Borel transform
of $\eta_V$ is located at $u=1$. From its residue, we obtain
\begin{equation}
   \Delta\eta_V = {C_F\over 2\beta_0}\,e^{-C}\,\Lambda_{\rm QCD}^2
   \,\bigg( {1\over m_1} - {1\over m_2} \bigg)^2 + O(N_f^{-2}) \,.
\end{equation}
Note that $\Delta\eta_V$ vanishes in the limit $m_1=m_2$, in which
the vector current is conserved and not renormalized, and hence
$\eta_V=1$.

As a second example, we demonstrate the cancellation of renormalon
ambiguities in the ratio of the vector form factor $h_V(w)$ defined
by
\begin{equation}
   \langle D^*(v_2,\epsilon)|\,\bar c\,\gamma^\mu b\,|\bar B(v_1)
   \rangle = \sqrt{m_B m_{D^*}}\,h_V(w)\,
   \epsilon^{\mu\nu\alpha\beta}\,\epsilon_\nu^*\,v_{2\alpha}
   v_{1\beta}
\end{equation}
and the axial form factor $h_{A_1}(w)$. Including power corrections
of order $1/m_c$ (and neglecting those of order $1/m_b$ and higher),
one finds that \cite{MN1}
\begin{equation}
   {h_V(w)\over h_{A_1}(w)} = {C_1^V(w)\over C_1^A(w)}\,\bigg\{
    1 + {\bar\Lambda\over m_c}\,\bigg[ {1\over w+1}
    + (w-1)\,{\partial\over\partial w}\,
    \ln{C_1^V(w)\over C_1^A(w)} \bigg] + \ldots \bigg\} \,.
\end{equation}
Using the first relation in (\ref{conrel}), we see that the infrared
renormalon of order $1/m_c$ in the ratio $C_1^V/C_1^A$ is precisely
compensated by the ultraviolet renormalon of the term proportional to
$\bar\Lambda$.

We conclude this section by pointing out an important implication of
our result (\ref{DetaA}). Recently, it has been claimed that one can
derive a sum rule for the form factor $h_{A_1}(1)$, from which it is
possible to obtain a bound for the non-perturbative corrections of
order $1/m_Q^2$ in (\ref{hA1}). The sum rule reads \cite{Bigisum}
\begin{equation}\label{sumrul}
   h_{A_1}^2(1) + \ldots = \eta_A^2 - {\lambda_2\over 3 m_c^2}
   + {\lambda_1+3\lambda_2\over 4}\,\bigg( {1\over m_c^2}
   + {1\over m_b^2} + {2\over 3 m_c m_b} \bigg) + O(1/m_Q^3) \,,
\end{equation}
where the ellipses represents positive contributions from transitions
into excited states. The parameters $\lambda_1$ and $\lambda_2$ are
defined in terms of the $B$-meson matrix elements of the kinetic and
the chromo-magnetic operator in the effective Lagrangian
(\ref{Lfull}). In principle, these HQET parameters could contain
ultraviolet renormalons. However, since $\lambda_2$ is proportional
to the mass splitting between $B$ and $B^*$ mesons, it is protected
from renormalons. Moreover, from the expansion of the meson mass
$m_B$ in powers of $1/m_b$ (this extends (\ref{Lamdef}) to order
$1/m_b$ \cite{FaNe})
\begin{equation}
   m_B = m_b^{\rm pole} + \bar\Lambda
   - {\lambda_1 + 3\lambda_2\over 2 m_b} + O(1/m_b^2) \,,
\end{equation}
and from the fact that to order $1/N_f$ the Borel transform of the
pole mass given in (\ref{mpoleu}) does not contain an infrared
renormalon pole at $u=1$, it follows that $\lambda_1$ does not
contain an ultraviolet renormalon\footnote{Note that even if
$\lambda_1$ and $\lambda_2$ would contain ultraviolet renormalons,
the mass dependence of the power corrections in (\ref{sumrul}) would
not match with the mass dependence of the infrared renormalon pole in
$\eta_A$ as given in (\ref{DetaA}).}
(at least) to order $1/N_f$ \cite{BBren}. We conclude that the
non-perturbative corrections in (\ref{sumrul}) do not contain the
ultraviolet renormalons required to cancel the infrared renormalon in
the perturbative coefficient $\eta_A^2$. Hence, there must be
something wrong with the sum rule. Either the short-distance
correction on the right-hand side is not given by $\eta_A^2$, or
there must be additional terms of order $1/m_Q^2$ to compensate the
renormalon in $\eta_A^2$. Therefore, the numerical implications
derived from this sum rule in Ref.~\cite{Bigisum} should be taken
with caution.

\section{Heavy-to-Light Transition Matrix Elements}
\label{sec:4}

We have seen in the previous section that the appearance of
renormalons in heavy-to-heavy transition matrix elements is to a
large extent constrained by the symmetries and equation of motion of
the HQET, which apply to both the initial and final hadron states. As
a consequence, ultraviolet renormalons enter the HQET matrix elements
of dimension-four operators only through the parameter $\bar\Lambda$
and a single function $K(w,\mu)$. Since the ultraviolet renormalon in
$\bar\Lambda$ is related to the infrared renormalon in the pole mass,
it is possible to derive the consistency relations (\ref{conrel}),
which determine the infrared renormalon poles in the coefficient
functions independently of any unknown hadronic matrix element.

It is well-known that in heavy-to-light transitions there are fewer
constraints imposed by heavy quark symmetry. In particular, most (if
not all) form factors appearing at order $1/m_Q$ mix with lower
dimensional operators under renormalization. Examples are provided by
the $1/m_Q$ expansions for meson decay constants \cite{sublea} and
the semileptonic $\bar B\to\pi\,\ell\,\bar\nu$ decay form factors
\cite{Btopi}. Therefore, it is not possible to derive consistency
relations analogous to (\ref{conrel}) in this case. The best one can
achieve is to deduce the structure of ultraviolet renormalons in the
hadronic form factors of the HQET from a calculation of the infrared
renormalons in the coefficient functions and the pole mass. We shall
discuss this for the simplest case of meson decay constants.

Consider heavy-to-light transition matrix elements of the form
$\langle X|\,\bar q\,\Gamma\,Q\,|H(v)\rangle$, where
$\Gamma=\gamma^\mu$ or $\gamma^\mu\gamma_5$, $H(v)$ is a heavy hadron
with velocity $v$, and $X$ is some light final state. For simplicity,
we set the mass of the light quark to zero and use a regularization
scheme with anticommuting $\gamma_5$. This leads to a simple relation
between the coefficient functions appearing in the expansion of the
vector and axial vector currents \cite{FNL}:
\begin{eqnarray}
   \bar q\,\gamma^\mu Q &\to& C_1(m_Q/\mu)\,\bar q\,\gamma^\mu h_v
    + C_2(m_Q/\mu)\,\bar q\,v^\mu h_v + O(1/m_Q)
    \,, \nonumber\\
   \phantom{ \bigg[ }
   \bar q\,\gamma^\mu\gamma_5\,Q &\to& C_1(m_Q/\mu)\,
    \bar q\,\gamma^\mu\gamma_5\,h_v - C_2(m_Q/\mu)\,
    \bar q\,v^\mu\gamma_5\,h_v + O(1/m_Q) \,.
\end{eqnarray}
These coefficients can be calculated in analogy to the previous
section. For their Borel transforms, we obtain
\begin{eqnarray}
   \widetilde C_1(u) &=& \delta(u) + {C_F\over\beta_0}\,\Bigg\{
    -3\,e^{-C u}\,\bigg( {\mu\over m_Q} \bigg)^{2u}\,\bigg( 1
    + {u\over 3} - u^2 \bigg)\,
    {\Gamma(u)\,\Gamma(1-2u)\over\Gamma(3-u)} \nonumber\\
   &&\phantom{ \delta(u) + {C_F\over\beta_0}\,\Bigg\{ }
    + {3\over 2 u} + R(u) \Bigg\} + O(N_f^{-2}) \,, \nonumber\\
   \widetilde C_2(u) &=& {4 C_F\over\beta_0}\,
    e^{-C u}\,\bigg( {\mu\over m_Q} \bigg)^{2u}\,
    {\Gamma(1+u)\,\Gamma(1-2u)\over\Gamma(3-u)} + O(N_f^{-2}) \,,
\end{eqnarray}
where $R(u)=5/4+O(u)$ in the $\overline{\rm MS}$ scheme. We have
checked that from an expansion around $u=0$ one recovers the known
one-loop expressions for the coefficient functions given in
Ref.~\cite{JiMu}. From the residues of the poles at $u=1/2$, it is
straightforward to compute the renormalon ambiguities
\begin{eqnarray}\label{DCihl}
   \Delta C_1 &=& -{11\over 12}\,{\Delta m_{\rm pole}\over m_Q}
    + O(N_f^{-2}) \,, \nonumber\\
   \Delta C_2 &=& {2\over 3}\,{\Delta m_{\rm pole}\over m_Q}
    + O(N_f^{-2}) \,.
\end{eqnarray}

To see how renormalons cancel in physical quantities, let us consider
the $1/m_Q$ expansion for pseudoscalar and vector meson decay
constants in the HQET. It reads \cite{sublea}
\begin{eqnarray}\label{fPfV}
   f_P\sqrt{m_P} &=& \Big[ C_1(m_Q/\mu) + C_2(m_Q/\mu) \Big]\,
    F(\mu)\,\Bigg\{ 1 + {1\over m_Q}\,\bigg[ G_1(\mu)
    - b(m_Q/\mu)\,{\bar\Lambda\over 6} \bigg] \nonumber\\
   &&\mbox{} + {6\over m_Q}\,\bigg[ C_{\rm mag}(m_Q/\mu)\,G_2(\mu)
    - B(m_Q/\mu)\,{\bar\Lambda\over 12} \bigg] \Bigg\}
    + O(1/m_Q^2) \,, \nonumber\\
   f_V\sqrt{m_V} &=& C_1(m_Q/\mu)\,
    F(\mu)\,\Bigg\{ 1 + {1\over m_Q}\,\bigg[ G_1(\mu)
    - b(m_Q/\mu)\,{\bar\Lambda\over 6} \bigg] \nonumber\\
   &&\mbox{} - {2\over m_Q}\,\bigg[ C_{\rm mag}(m_Q/\mu)\,G_2(\mu)
    - B(m_Q/\mu)\,{\bar\Lambda\over 12} \bigg] \Bigg\}
    + O(1/m_Q^2) \,,
\end{eqnarray}
where $C_{\rm mag}(m_Q/\mu) = 1+O(\alpha_s)$ is the coefficient of
the chromo-magnetic operator in the effective Lagrangian
(\ref{Lfull}), while $B(m_Q/\mu) = 1+O(\alpha_s)$ and $b(m_Q/\mu) =
O(\alpha_s)$ are coefficients that appear at order $1/m_Q$ in the
expansion of the currents \cite{sublea}. $F(\mu), G_1(\mu)$, and
$G_2(\mu)$ are hadronic parameters, which are independent of $m_Q$.
Both $G_1(\mu)$ and $G_2(\mu)$ mix with lower dimensional operators
and contain ultraviolet renormalons, as can be seen from the
renormalization-group equations \cite{sublea}
\begin{eqnarray}
   \mu\,{{\rm d}\over{\rm d}\mu}\,G_1(\mu)
   &=& {\bar\Lambda\over 6}\,
    \mu\,{{\rm d}\over{\rm d}\mu}\,b(m_Q/\mu) \,, \nonumber\\
   \mu\,{{\rm d}\over{\rm d}\mu}\,\Big[ C_{\rm mag}(m_Q/\mu)\,
   G_2(\mu) \Big] &=& {\bar\Lambda\over 12}\,
    \mu\,{{\rm d}\over{\rm d}\mu}\,B(m_Q/\mu) \,.
\end{eqnarray}
Requiring that in (\ref{fPfV}) the infrared renormalons in the
coefficient functions cancel against the ultraviolet renormalons in
$\bar\Lambda$ and $G_i(\mu)$, we obtain the relations
\begin{eqnarray}
   {\Delta C_1 + \Delta C_2\over C_1 + C_2}
   &=& - {\Delta m_{\rm pole}\over 6 m_Q}\,(b+3B)
    - {1\over m_Q}\,\Big( \Delta G_1
    + 6 C_{\rm mag}\,\Delta G_2 \Big) \,, \nonumber\\
   {\Delta C_1\over C_1} &=& - {\Delta m_{\rm pole}\over 6 m_Q}\,
    (b-B) - {1\over m_Q}\,\Big( \Delta G_1 - 2 C_{\rm mag}\,
    \Delta G_2 \Big) \,.
\end{eqnarray}
To order $1/N_f$, they simplify to
\begin{eqnarray}
   \Delta G_1 &=& - {m_Q\over 4}\,\Big( 4\Delta C_1
    + \Delta C_2 \Big) + O(N_f^{-2})
    = {3\Delta m_{\rm pole}\over 4} + O(N_f^{-2}) \,, \nonumber\\
   \Delta G_2 &=& - {\Delta m_{\rm pole}\over 12}
    - {m_Q\over 8}\,\Delta C_2 + O(N_f^{-2})
    = - {\Delta m_{\rm pole}\over 6} + O(N_f^{-2}) \,.
\end{eqnarray}
This determines the ultraviolet renormalon ambiguities in the
hadronic parameters $G_i(\mu)$. The situation encountered here is
general for heavy-to-light transitions; since there are always at
least two hadronic parameters that contain ultraviolet renormalons,
it is not possible to derive a consistency condition for the infrared
renormalons in the coefficient functions $C_1$ and $C_2$. However,
the residues of the renormalons in the coefficient functions
determine in a unique way the residues of the ultraviolet renormalon
poles in the hadronic parameters of the HQET.

\section{Inclusive Decays of Heavy Hadrons}
\label{sec:5}

\begin{figure}[h]
   \vspace{-3cm}
   \epsfxsize=11cm
   \centerline{\epsffile{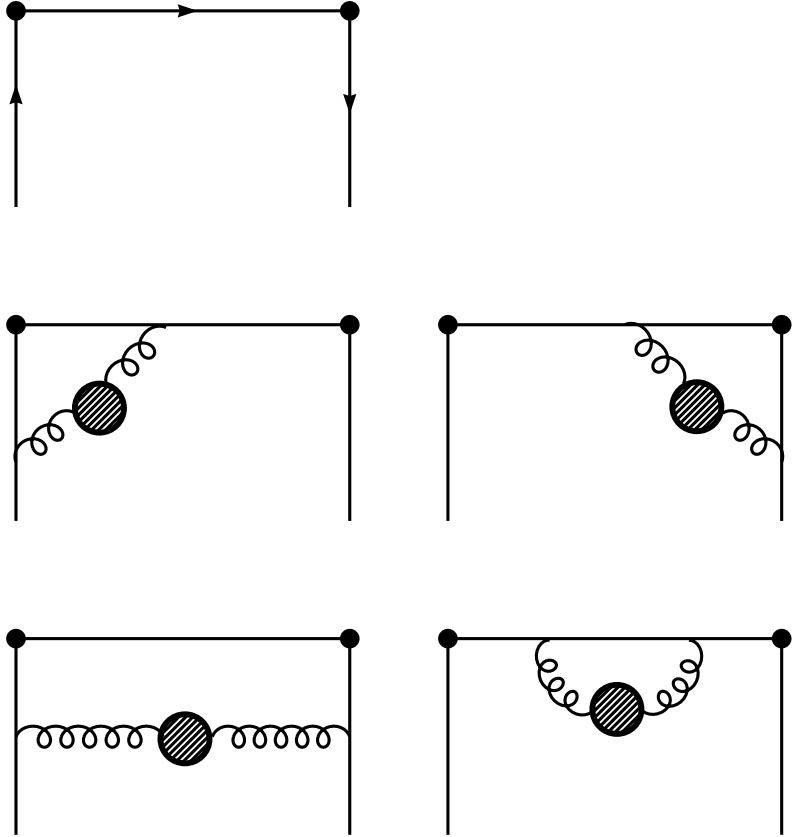}}
   \centerline{\parbox{13cm}{\caption{\label{fig:3}
Tree-level contribution and radiative corrections to the transition
amplitude.}}}
\end{figure}

After the analysis of exclusive transitions, we will now consider
current-induced inclusive decays of hadrons containing a heavy quark.
Examples are the semileptonic decays $\bar B\to X_q\,\ell\,\bar\nu$
and $\Lambda_b\to X_q\,\ell\,\bar\nu$, where $q=c$ or $u$, as well as
the rare decay $\bar B\to X_s\,\gamma$. The flavour-changing current
relevant to semileptonic decays is $\bar q\,\gamma^\mu(1-\gamma_5)\,
b$. For the penguin-induced transitions, it is of the form $\bar
q\,\sigma^{\mu\nu} (1\pm\gamma_5)\,b$. The inclusive decay
distributions can be calculated in powers of $1/m_b$ using an OPE for
the transition amplitude \cite{Chay}--\cite{shape}
\begin{equation}
   T(v,p) = -i\int{\rm d}^4 x\,e^{-ip\cdot x}\,
   \langle H(v)|\,{\rm T}\,\Big\{ \bar b(x)\,\Gamma_1\,q(x),
   \bar q(0)\,\Gamma_2\,b(0) \Big\}\,| H(v)\rangle \,,
\end{equation}
where $H(v)$ denotes the decaying $b$-flavoured hadron with velocity
$v$, $p$ is the momentum carried by the current (in the cases above,
the total lepton or photon momentum, respectively), and $\Gamma_i$
are abbreviations for the appropriate Dirac matrices.
 The OPE is constructed by performing a phase redefinition
[cf.~(\ref{redef})]
\begin{equation}\label{bvdef}
   b_v(x) =  \exp(-i m_b\,v\cdot x)\,b(x) = h_v(x) + O(1/m_b)
\end{equation}
to pull out the leading dependence of the fields on the heavy quark
mass. The next step is to write $T(v,p)$ as a sum of coefficient
functions multiplying local, higher dimensional operators. The
coefficients are determined by evaluating the diagrams shown in
Fig.~\ref{fig:3}, where the momentum of the $b$-quark is as usual
written in the form $p_b=m_b\,v+k$. The residual momentum $k$ is
equivalent to a derivative acting on the rescaled heavy quark field
$b_v$.

We will evaluate the contributions in the OPE including terms of
order $1/m_b$ and $1/N_f$. In general, the equation of motion can be
used to relate all terms of order $1/m_b$ to the residual mass term
in the HQET Lagrangian, which is itself of order $1/N_f$. Hence, it
will be sufficient to evaluate the $1/m_b$ corrections at tree level.
Let us then start with the discussion of the tree diagram in
Fig.~\ref{fig:3}. It gives
\begin{equation}
   T_{\rm tree} = \langle\,\Gamma_1\,
   {1\over m_b\,\rlap/v - \rlap{\,/}p - m_q + i\rlap{\,/}D}\,
   \Gamma_2\,\rangle \,,
\end{equation}
where we use the short-hand notation $\langle H(v)|\,\bar b_v\,
\Gamma\,b_v\,|H(v)\rangle\equiv\langle\,\Gamma\,\rangle$. The tree
diagram contains the propagator of the $q$-quark in the background
field of the light degrees of freedom in the decaying hadron. To
proceed, we expand the propagator as
\begin{equation}\label{qprop}
   {1\over m_b\,\rlap/v - \rlap{\,/}p - m_q + i\rlap{\,/}D}
   = {1\over m_b\,\rlap/v - \rlap{\,/}p - m_q}
   - {1\over m_b\,\rlap/v - \rlap{\,/}p - m_q}\,i\rlap{\,/}D\,
   {1\over m_b\,\rlap/v - \rlap{\,/}p - m_q} + \ldots \,,
\end{equation}
where the second term is of order $1/m_b$ relative to the first term.
The forward matrix element of any local operator $\bar
b_v\,\Gamma_\alpha\,i D^\alpha\,b_v$ containing a single covariant
derivative can be evaluated, up to $1/m_b$ corrections, using
(\ref{bvdef}) together with the equation of motion $i v\cdot D\,h_v =
\delta m\,h_v$, where $\delta m$ is the residual mass term. It
follows that
\begin{equation}
   \langle\,\Gamma_\alpha\,i D^\alpha\,\rangle = \delta m\,
   \langle\,\Gamma_\alpha\,v^\alpha\,\rangle + \ldots \,,
\end{equation}
where $\Gamma_\alpha$ denotes an arbitrary Dirac matrix, and the
ellipses represent terms that are suppressed by one power of $1/m_b$.
Applying this relation, and resumming the expanded propagator
(\ref{qprop}), we find
\begin{equation}
   T_{\rm tree} = \langle\,\Gamma_1\,
   {1\over (m_b+\delta m)\,\rlap/v - \rlap{\,/}p - m_q}\,
   \Gamma_2\,\rangle + \ldots \,,
\end{equation}
where the ellipses represent terms of order $1/m_b^2$ relative to the
leading term. We observe that, as in the case of exclusive decays,
the residual mass term always appears together with the HQET
expansion parameter $m_b$ in the combination $m_b^{\rm pole} =
m_b+\delta m$, i.e.\ it is the pole mass that enters the tree-level
expression for the transition amplitude. The infrared renormalon in
the pole mass leads to an ambiguity given by
\begin{equation}
   \Delta T_{\rm tree} = \langle\,\Gamma_1\,
   {1\over m_b\,\rlap/v - \rlap{\,/}p - m_q}\,
   (-\Delta m_{\rm pole}\,\rlap/v)\,
   {1\over m_b\,\rlap/v - \rlap{\,/}p - m_q}\,\Gamma_2\,\rangle \,.
\end{equation}

To see how this renormalon is cancelled, let us now turn to the
calculation of the radiative corrections depicted in
Fig.~\ref{fig:3}. We study the Borel transform of the transition
amplitude to order $1/N_f$ using the resummed gluon propagator
(\ref{Dab}). In the calculation, we only keep terms that have a
renormalon pole at $u=1/2$. We obtain:
\begin{eqnarray}
   \widetilde T_{\rm vertex}(u) &=& {8 C_F\over\beta_0}\,
    e^{-C/2}\,\mu\,\Gamma(1-2u)\,
    \langle \Gamma_1 {1\over m_b\,\rlap/v - \rlap{\,/}p - m_q}\,
    \rlap/v\,{1\over m_b\,\rlap/v - \rlap{\,/}p - m_q}\,\Gamma_2
    \rangle + \ldots , \nonumber\\
   \widetilde T_{\rm box}(u) &=& - {4 C_F\over\beta_0}\,
    e^{-C/2}\,\mu\,\Gamma(1-2u)\,
    \langle\,\Gamma_1\,{1\over m_b\,\rlap/v - \rlap{\,/}p - m_q}\,
    \rlap/v\,{1\over m_b\,\rlap/v - \rlap{\,/}p - m_q}\,\Gamma_2\,
    \rangle \nonumber\\
   &&\mbox{}+ {6 C_F\over\beta_0}\,
    e^{-C/2}\,{\mu\over m_b}\,\Gamma(1-2u)\,
    \langle\,\Gamma_1\,{1\over m_b\,\rlap/v - \rlap{\,/}p - m_q}\,
    \Gamma_2\,\rangle + \ldots \,, \nonumber\\
   \widetilde T_{\rm WFR}(u) &=& - {6 C_F\over\beta_0}\,
    e^{-C/2}\,{\mu\over m_b}\,\Gamma(1-2u)\,
    \langle\,\Gamma_1\,{1\over m_b\,\rlap/v - \rlap{\,/}p - m_q}\,
    \Gamma_2\,\rangle + \ldots \,.
\end{eqnarray}
The ellipses represent terms that are regular at $u=1/2$, and terms
of order $1/N_f^2$. Note that there is no renormalon contribution
from the renormalization of the $q$-quark propagator. Moreover, the
renormalon poles with residues proportional to the tree diagram
cancel between the box graph and wave-function renormalization. For
the sum of all loop contributions, we find
\begin{equation}
   \widetilde T_{\rm loops}(u) = {4 C_F\over\beta_0}\,
   e^{-C/2}\,\mu\,\Gamma(1-2u)\,
   \langle\,\Gamma_1\,{1\over m_b\,\rlap/v - \rlap{\,/}p - m_q}\,
   \rlap/v\,{1\over m_b\,\rlap/v - \rlap{\,/}p - m_q}\,\Gamma_2\,
   \rangle + \ldots \,.
\end{equation}
{}From the residue of the pole at $u=1/2$, we obtain for the
renormalon ambiguity
\begin{equation}
   \Delta T_{\rm loops} = \langle\,\Gamma_1\,
   {1\over m_b\,\rlap/v - \rlap{\,/}p - m_q}\,
   \Delta m_{\rm pole}\,\rlap/v\,
   {1\over m_b\,\rlap/v - \rlap{\,/}p - m_q}\,\Gamma_2\,\rangle
   + O(N_f^{-2}) \,.
\end{equation}
As expected, the sum of all contributions in the OPE for the
transition amplitude is free of renormalon ambiguities:
\begin{equation}
   \Delta T = \Delta T_{\rm tree} + \Delta T_{\rm loops} = 0 \,.
\end{equation}
That this cancellation occurs was conjectured by Bigi et al.\ in
Ref.~\cite{Bigiren}, however without presenting an explicit
calculation. In fact, it was claimed that infrared renormalons only
appear in the vertex corrections and mass renormalization, but not in
the box diagram. Our calculation shows that this is not
correct.\footnote{We note that the renormalon contributions in the
individual diagrams are the same in all covariant gauges.}
Nevertheless, we confirm that the cancellation occurs when all
diagrams are taken into account.

The situation encountered here is special in that to order $1/m_b$
there do not appear non-perturbative corrections when the pole mass
is used in the OPE of the transition amplitude. Hence, at this order
there are no ultraviolet renormalons. What we have demonstrated above
is a cancellation of infrared renormalons. Consider, as an example,
the total decay rate for the process $\bar B\to X_u\,\ell\,\bar\nu$.
It can be calculated from the imaginary part of the transition
amplitude. Neglecting the mass of the $u$-quark, one obtains the
well-known result \cite{Burate}
\begin{equation}\label{rate}
   \Gamma(\bar B\to X_u\,\ell\,\bar\nu)
   = {G_F^2\,|\,V_{ub}|^2\over 192\pi^3}\,C(m_b)\,
   \Big\{ 1 + O(1/m_b^2) \Big\} \,,
\end{equation}
where
\begin{equation}
   C(m_b) = \big( m_b^{\rm pole} \big)^5\,\bigg\{ 1
   - {2\alpha_s(m_b)\over 3\pi}\,\bigg( \pi^2 - {25\over 4}
   \bigg) + \ldots \bigg\} \,.
\end{equation}
We have shown that the infrared renormalon at $u=1/2$ in the pole
mass is cancelled by an infrared renormalon in the perturbative
series. It is possible to eliminate these renormalons explicitly by
introducing a heavy quark mass $m_b^{\rm R}$ renormalized at short
distances instead of using the pole mass \cite{Bigiren,Patr}. As long
as $m_b^{\rm R}$ differs from $m_b^{\rm pole}$ by a multiplicative
factor $Z[\alpha_s(m_b)]$, this substitution does not induce $1/m_b$
corrections to the decay rate (\ref{rate}). For instance, we may work
in the $\overline{\rm MS}$ scheme and use the running mass
$\overline{m}_b(\mu)$ evaluated at $\mu=m_b$. This leads to
\begin{equation}
   C(m_b) = \big[ \overline{m}_b(m_b) \big]^5\,\bigg\{ 1
   - {2\alpha_s(m_b)\over 3\pi}\,\bigg( \pi^2 - {65\over 4}
   \bigg) + \ldots \bigg\} \,.
\end{equation}
The perturbative series in this expression does no longer contain a
renormalon at $u=1/2$. Note, however, that at some higher order in
the $1/m_b$ expansion there will appear ultraviolet renormalons in
the non-perturbative corrections to the decay rate (\ref{rate}).
Correspondingly, the coefficient $C(m_b)$ must contain infrared
renormalons at larger values of the Borel parameter $u$, which cannot
be eliminated by introducing the renormalized mass $m_b^{\rm R}$.

\section{Summary and Conclusions}
\label{sec:6}

We have investigated the appearance of renormalons in the HQET by
considering the $1/m_Q$ expansion for exclusive heavy-to-heavy and
heavy-to-light transitions, as well as for inclusive decays of heavy
hadrons. We have argued that, in general, infrared renormalons in the
coefficient functions of HQET operators are compensated by
ultraviolet renormalons in the matrix elements of higher dimensional
operators, and we have identified which of the HQET matrix elements
contain such ultraviolet renormalons. In the case of heavy-to-heavy
transitions, the symmetries and the equation of motion of the
effective theory lead to five consistency relations among the
infrared renormalons in the pole mass and the coefficient functions.
We have checked that these relations are satisfied to next-to-leading
order in an expansion in powers of $1/N_f$.

The most important, though not surprising, result of our analysis is
that the appearance of renormalons does not alter the structure of
the heavy quark expansion, and does not invalidate any of the
predictions derived using the HQET. In particular, Luke's theorem, as
well as relations between weak decay form factors, remain valid. In
this sense, there is no ``renormalon problem'' in the HQET. However,
as in any OPE it is true that some of the dimensionful hadronic
parameters describing the non-perturbative corrections in the heavy
quark expansion have an intrinsic uncertainty of order $\Lambda_{\rm
QCD}^n$. An example is provided by the mass parameter $\bar\Lambda$.
In the practical form of the OPE, in which dimensional regularization
is employed in the calculation of the coefficient functions,
ambiguities arise from the necessity to specify a resummation
prescription to regulate the divergent asymptotic behaviour of
perturbation theory. In the literal form of Wilson's OPE, they arise
from the introduction of a hard factorization scale $\mu$. The
hadronic parameters of the effective theory then exhibit a power-like
dependence on $\mu$, in a way that depends on how the cutoff is
implemented. In both cases, to define these parameters precisely
would require one to fix terms in the coefficient functions that are
exponentially small in the coupling constant. As long as one works
with truncated perturbative expressions for the Wilson coefficients,
the errors due to the truncation are parametrically larger than power
corrections. This type of ambiguity is inherent in any OPE and as
such cannot be avoided. In this context, we note that the
introduction of a short-distance mass instead of the pole mass, which
was proposed in Refs.~\cite{Bigiren,Patr}, does in general not help
to eliminate renormalons. An exception is the case of inclusive
decays of heavy hadrons, where this procedure eliminates the leading
infrared renormalons. On the other hand, such a choice of the heavy
quark mass destroys the flavour symmetry of the effective Lagrangian
of the HQET and is thus unattractive, at least in processes that
involve more than one heavy quark flavour.

Finally, we like to point out that our somewhat formal investigation
of renormalons can serve for tests of HQET calculations. In some
cases, the requirement that a compensation of infrared and
ultraviolet renormalons occurs leads to non-trivial relations. An
example is provided by the consistency conditions (\ref{conrel}) for
heavy-to-heavy transitions. Using a similar argument, we could show
with an explicit calculation that a sum rule derived by Shifman et
al.\ \cite{Bigisum}, which has been used to put a bound on the
hadronic form factor that enters the extraction of $|\,V_{cb}|$ from
semileptonic decays, must be incorrect. This sum rule relates a
physical observable to a theoretical expression in which infrared
renormalons in a coefficient function do not match with ultraviolet
renormalons in non-perturbative parameters. This expression has an
intrinsic ambiguity and thus cannot be complete. A further
investigation of what goes wrong with the argument presented in
Ref.~\cite{Bigisum} is necessary before any useful phenomenological
bound can be derived.

\bigskip
While this paper was in writing, we became aware of a preprint by
Beneke et al.\ \cite{BBZ}, who demonstrate the cancellation of
infrared renormalons in inclusive decay rates. Their results agree
with our Sect.~\ref{sec:5}.

\subsection*{Acknowledgements}

It is a pleasure to thank Martin Beneke, Vladimir Braun and Guido
Martinelli for interesting discussions. CTS thanks Prof.\ J. Ellis
and members of the theory division at CERN for their hospitality
during the early stages of this work, and acknowledges the Particle
Physics and Astronomy Research Council for their support through the
award of a Senior Fellowship.

\newpage

\centerline{\Large\bf Figures}

\vfil

\end{document}